\documentclass[final]{Dekker}
\usepackage{amsmath}
\usepackage{amssymb}
\begin{document}
\setcounter{page}{1}
\jname{TRANSPORT THEORY AND STATISTICAL PHYSICS} 
\jvol{ }
\jissue{ }
\jyear{ }
\webslug{www.dekker.com}
\cpright{Marcel Dekker, Inc.}
\title[CAUCHY INTEGRAL EQUATIONS. II]{RADIATIVE TRANSFER IN PLANE-PARALLEL MEDIA AND CAUCHY INTEGRAL EQUATIONS II. THE $H$-FUNCTION}
\author[RUTILY, BERGEAT AND CHEVALLIER]{B. Rutily, J. Bergeat, L. Chevallier
\affiliation{Centre de Recherche Astronomique de Lyon (UMR 5574 du CNRS),
Observatoire de Lyon,\\
9, Avenue Charles Andr\'e,
69561 Saint-Genis-Laval Cedex, France\\ \ \\ Submitted January 2003}}

%\date{January 2003}

\abstract{In the central part of this paper, we revisit the classical study of the $H$-function defined as the unique solution, regular in the right complex half-plane, of a Cauchy integral equation. We take advantage of our work on the $N$-function published in the first article of this series. The $H$-function is then used to solve a class of Cauchy integral equations occurring in transfer problems posed in plane-parallel media. We obtain a concise expression of the unique solution analytic in the right complex half-plane, then modified with the help of the residue theorem for numerical calculations.}

\keywords{Radiative transfer equation; Plane-parallel geometry; Isotropic scattering; Cauchy integral equations}

\maketitle

\section{\label{Sec1} INTRODUCTION}
This article is the second in a series dedicated to the Cauchy integral equations arising in radiative transfer problems in plane-parallel geometry. We adopted in Ref. [1] (hereafter I) the simplest scattering law--monochromatic and isotropic--with volumic albedo $0<a<1$ and assumed that light propagates through a homogeneous and stationnary slab of optical thickness $0<b\leq+\infty$. The resulting integral equations take the form
\begin{equation}
T(a,z)f(z)-\frac{a}{2}z\int_{0}^{1}f(v)\frac{dv}{v-z} = c(z),
\end{equation}
where $T(a,z)$ is the dispersion function defined for any $z\in\mathbb{C}\setminus\lbrace\pm1\rbrace$ by
\begin{equation}
T(a,z) = 1+\frac{a}{2}z\int_{-1}^{+1}\frac{dv}{v-z}.
\end{equation} 

The integrals in Eqs. (1) and (2) are Cauchy principal values (symbol $f$) for $z\in]0,1[$ and $z\in]-1,+1[$ respectively. Thus, Eq. (1) is both a Cauchy integral equation over $[0,1]$ and the definition of the extension of $f$ outside this interval.

Given a function defined by a Cauchy-type integral, the same symbol is used for the values of the function outside the cut and on the cut. This writing convention allows to extend, on the cuts, the relations satisfied by those functions which include Cauchy-type integrals in their definition. For instance the definition (2) of the $T$-function becomes, for any $z = u\in]-1,+1[$
\begin{equation}
T(a,u) = 1+\frac{a}{2}u \overset{+1}{\underset{-1}{f}} \frac{dv}{v-u} = 1-\frac {a}{2}u\ln[\frac{1+u}{1-u}],
\end{equation}
where $u$ denotes the $z$-variable while exploring the real axis.

The right-hand side $c(z)$ of Eq. (1) is a given function satisfying the following properties:

({\it i}) it is defined in $\mathbb{C}$ except possibly at $z = 0$ and, for finite media, at $z = -1$,

({\it ii}) it is analytic in the half-plane $\Re(z)\geq0$, except possibly at $z = 0$ $[\Re(z)$ = real part of $z\in\mathbb{C}]$,

({\it iii}) it is bounded on the right at $z = 0$, i.e., the limit of $c(z)$ when $z$ tends to 0 with $\Re(z)>0$ exists and is finite.
     
The space of solutions to Eq. (1) at least H\"{o}lder continuous on the line segment $[0,1]$ is one-dimensional. The solutions can be written in the form (I, 44) in $\mathbb{C}\setminus[0,1]$, and in the form (I, 19) in $]0,1[$, where $P_0$ is a constant [Eq. ($n$) of I is referred to as (I, $n$)]. End-points of the segment $[0,1]$ and points where the function $z\rightarrow T(a,z)$ vanishes or diverges are particular cases for Eq. (1), that are discussed in Sec. 4 of I. Specifically, Eqs. (I, 45) and (I, 46) are satisfied at $z = 0$, and Eq. (I, 49) at $z = +1$.

In the present paper, we only deal with the solution to Eq. (1) which is analytic in the right complex half-plane. As shown by Mul\-likin,$^{[2,3]}$ this condition determines the constant $P_0$ in Eqs. (I, 44), (I, 19), (I, 49), and guarantees the uniqueness of the solution. The analyticity condition, in the right half-plane, for the solution to Eq. (1), is a fundamental requirement of the transfer theory in plane-parallel media. It follows from the fact that the auxiliary functions of this theory, which satisfy equations of the form (1), are derived through the Laplace transform over $[0,b]$ of the Green distribution of the problem.$^{[4]}$ In a half-space $(b = +\infty)$, this Laplace transform exists and is analytic at least in the right complex half-plane. In a finite slab $(b < +\infty)$, this property is satisfied in the whole complex plane.$^{[4]}$

Once the value of the constant $P_0$ which leads to the required solution is derived (Sec. \ref{Sec2}), we express this solution for the specific free term $c(z) = 1$ (Sec. \ref{Sec3}). This is the well known $H$-function for semi-infinite media, which is of particular interest for the theory of integral equations of the form (1). Its relation to the $N$-function is further clarified and its analytical properties in the whole complex plane are deduced {\it in a self-consistent manner} from those for the $N$-function.

While the $N$-function characterizes the set of solutions to Eq. (1), the $H$-function expresses its unique solution analytic in the right complex half-plane. It is therefore natural to write this solution in terms of $H$, which is done in Sec. \ref{Sec4}. The expression (50) for this solution is the basis for the calculation of the $X$- and $Y$-functions in finite media (to be dealt with in the third paper of the present series).

\section{\label{Sec2}THE GENERAL CASE}
Equation (1), when solved in the interval $]0,1[$, defines the $f$-function in the whole complex plane. If we impose this function to be analytic in the right complex half-plane, we must have in particular $f(+1/k)<+\infty$, where $k = k(a)$ is the unique root in $]0,1[$ of the transcendental equation $T(a,1/z) = 0$, i.e.,
\begin{equation}
T(a,1/k)=1-\frac{a}{2}\frac{1}{k}\ln\frac{1+k}{1-k} = 0 \quad (0<k<1). 
\end{equation}

From Eq. (1), this forces the condition
\begin{equation}
\frac{a}{2}\int_{0}^{1}f(v)\frac{dv}{1-kv}=c(+1/k). 
\end{equation}

The constant $P_0$ in the solution (I, 44), (I, 19), (I, 49) is therefore constrained to a unique value.$^{[2,3]}$ To derive this value, multiply both members of Eq. (I, 19) by $(a/2)/(1-ku)$, integrate with respect to $u$ between 0 and 1, and use Eqs. (5), (I, 30) and (I, 32) with $z = +1/k$. The result is
\begin{equation}
P_{0}=-\frac{2}{a}\frac{k c(+1/k)}{N(a,+1/k)} + k\int_{0}^{1}(\gamma/N)(a,v)c(v)\frac{dv}{1-kv},  
\end{equation}
where $(\gamma/N)(a,v) = \gamma(a,v)/N(a,v)$, a notation we adopt hereafter, 
\begin{equation}
\gamma(a,v)=\frac{1}{{[T^{2}(a,v)+{(\pi\frac{a}{2}v)}^{2}]}^{1/2}}\quad(0\leq v <+1),  
\end{equation}
and for $z\in\mathbb{C}\setminus\lbrace+1\rbrace$
\begin{equation}
N(a,z)=\frac{1}{1-z}\exp\left\lbrace\frac{1}{\pi}\int_{0}^{1} \arctan[\frac{\pi\frac{a}{2}v}{T(a,v)}]\frac{dv}{v-z}\right\rbrace.  
\end{equation}

Entering the expression (6) into Eqs. (I, 44), (I, 19) and (I, 49) leads to the following solution to Eq. (1), analytic in the right complex half-plane:

\noindent
- for $z\in\mathbb{C}\setminus\lbrace[0,1]\cup\lbrace+1/k\rbrace \rbrace$
\begin{eqnarray}
\lefteqn{f(z)=\frac {c(z)}{T(a,z)}+z\frac{N(a,z)}{T(a,z)}\left\lbrace-\frac{kc(+1/k)}{N(a,+1/k)}\right.}\nonumber\\&&\qquad\qquad\quad\left.+\frac{a}{2}\int_{0}^{1}(\gamma/N)(a,v)c(v)[\frac{k}{1-kv}+\frac {1}{v-z}]dv \right\rbrace,  
\end{eqnarray}
- for $u\in]0,1[$ 
\begin{eqnarray}
\lefteqn{f(u)=(\gamma^{2}T)(a,u)c(u)+u(\gamma N)(a,u) \left\lbrace -\frac{kc(+ 1 / k)}{N(a,+1/k)}\right.}\nonumber\\&&\qquad\quad\qquad\left.+\frac{a}{2}\overset{1}{\underset{0}{f}}(\gamma/N)(a,v)c(v)[\frac{k}{1-kv}+\frac{1}{v-u}] dv \right\rbrace, 
\end{eqnarray}
- at $u = 0$
\begin{equation}
f(+0)=c(+0)<+\infty,\end{equation}
- and at $u = +1$
\begin{eqnarray}
\lefteqn{f(+1) = \frac{1}{1-a}\frac{k^2}{1-k^2}\frac{1}{N(a,-1)}\left\lbrace - \frac{kc(+1/k)}{N(a,+1/k)}\right.}\nonumber\\&&\qquad\qquad\quad\left.+\frac{a}{2} \int_ {0}^{1}(\gamma/N)(a,v)c(v)[\frac {k}{1-kv} + \frac {1}{v-1}] dv \right\rbrace. 
\end{eqnarray}

Note that Eq. (9) is not valid at those points where the function $T(a,z)$ vanishes or diverges. The first ones are $\pm 1/k$ and $-1/K$, where $1/k$ ($0<k<1$) and $1/K$ ($K>1$) are the positive roots of the equation $T(a,z) = 0$. At $u = +1/k$, the solution is bounded and equation (1) is indeterminate. From Eq. (9), we remove the indetermination by continuity, which gives
\begin{eqnarray}
\lefteqn{f(+1\!/k)=R(a,k)N(a,\!+1\!/k)\!\left\lbrace\frac{1}{k}\frac{c^{\prime}(+1\!/k)}{N(a,\!+1\!/k)}-\frac{1}{k}{\sqrt{1\!-\!a}}\;c(+1\!/k)\right.}\nonumber\\&&\qquad\qquad\left.+\frac{a}{2}\int_{0}^{1}\!(\gamma/N)(a,v)[c(v)-c(+1/k)]\frac{dv}{(1\!-\!kv)^2}\right\rbrace. 
\end{eqnarray}
In this expression, $c'(+1/k)$ is the derivative of $c$ at $+1/k$ ; it exists since the $c$-function is analytic in the right complex half-plane. We have set 
\begin{equation}
R(a,k)=\frac{1-k^2}{k^2+a-1}. 
\end{equation}

At the points $u = -1/k$ and $u = -1/K$, the solution necessarily diverges in a semi-infinite medium, as observed in a remark following Eq. (I, 47). We thus have
\begin{equation}
b=+\infty \Rightarrow f(-1/k)=f(-1/K)=\infty.  
\end{equation}

At $u = -1$, the function $T(a,u)$ diverges and the solution is always bounded (see I, Sec. 4). This solution is zero if, and only if, the $c$-function is bounded at -1, which happens in semi-infinite media
\begin{equation}
b=+\infty \Rightarrow f(-1)=0.  
\end{equation}

Equations (9)-(13) can be simplified for specific free terms $c(z)$, using the integral properties of the function  $\gamma/N$ on $[0, 1[$ as established in I. This is the case specially when $c(z) = 1$: from Eqs. (I, 32), (I, 33) and (I, 43), we derive the integrals appearing on the right-hand side of the solution (9)-(13), which then coincides with the $H$-function.

\section{\label{Sec3} THE $c = 1$ CASE: INTRODUCTION OF THE $H$-FUNCTION}
The $H$-function can be defined as the unique solution, analytic in the right complex half-plane, to Eq. (1) for the free term $c(z) = 1$. It satisfies the following integral equation, the so-called $H$-singular equation
\begin{equation}
T(a,z)H(a,z)=1+\frac{a}{2}z\int_{0}^{1}H(a,v)\frac{dv}{v-z}.  
\end{equation}

This equation is of Cauchy type over the line segment [0, 1] and has a unique solution satisfying the constraint (5), which becomes here
\begin{equation}
\frac {a}{2}\int_0^1H(a,v)\frac{dv}{1-kv}=1.  
\end{equation}

The solution of Eqs. (17)-(18) is based on the properties of the $N$-function we synthesized in I. This study is also at the root of the classical relations satisfied by the $H$-function, as solely deduced from its definition (17)-(18). We shall return to these two points in the next subsections.

\subsection{\label{Subsec. 3.A} Solution to the problem (17)-(18): Mullikin's expressions of the $H$-function}
We can write this solution by replacing $c(z)$ by 1 in the expressions (9)-(13), and evaluating the integrals with the help of Eqs. (I, 32), (I, 33) and (I, 43). We obtain:

\noindent
- for $z\in\mathbb{C}\setminus\lbrace\,]0,1]\cup\lbrace-1/k,-1,-1/K,+1/k\rbrace \rbrace$
\begin{equation}
H(a, z)=\frac{1}{k}\sqrt{1-a}\,(1-kz)(N/T)(a,z),  
\end{equation}
- for $u \in [0,1[$
\begin{equation}
H(a,u)= \frac{1}{k}\sqrt{1-a}\,(1-ku)(\gamma N)(a,u),  
\end{equation}
- at $u = +1$
\begin{equation}
H(a,+1)= \frac{k}{\sqrt{1-a}}\frac{1}{1+k}\frac{1}{N(a,-1)},  
\end{equation}
- and at $u = +1/k$
\begin{eqnarray}
H(a,+1/k)&=&-\frac{1}{k}{\sqrt{1-a}}\,R(a,k)N(a,+1/k),\\&=&\frac{1}{2}\frac{k}{\sqrt{1-a}}\frac{1}{N(a,-1/k)}, 
\end{eqnarray}
owing to Eq. (I, 36) [$R(a,k)$ as given by Eq. (14)]. Equations (19) and (20) are still valid at 0, and they yield $H(a,0) = 1$ according to Eq. (I, 39). This result is consistent with Eq. (I, 45). Equation (19) implies that the $H$-function diverges at $-1/k$ and $-1/K$, and tends to zero as $z$ tends to -1, which agrees with relations (15) and (16). The $H$-function can thus be extended by continuity at $u=-1$ and we have
\begin{equation}
H(a,-1/k) = H(a,-1/K) = \infty \;\; {\rm and} \;\; H(a,-1) = 0.
\end{equation}

Finally, putting $z \rightarrow \infty $ in Eq. (19) and using Eq. (I, 42) together with the relation $T(a,\infty) = 1-a$, we obtain $ H(a,\infty) = 1/\sqrt{1-a}$.

The expression (20) for the $H$-function is not suitable for numerical evaluation, since it contains a Cauchy integral in the $N(a,u)$-term. This difficulty vanishes when replacing $(\gamma N)(a,u)$ by its expression from the factorization relation (I, 27), which is valid here since $u\in[0, 1[$. Relation (19) may be treated in a similar way, replacing $N/T$ by its expression as given by Eq. (I, 26) for $z\in \mathbb{C}\setminus[-1,+1]$. The two expressions of $H$ thereby derived coincide, which proves that the result is valid for $z\in\mathbb{C}\setminus[-1,0\,[$. It can be written as
\begin{equation}
H(a,z)=\frac{k}{\sqrt{1-a}}\frac{1}{1+kz}\frac{1}{N(a,-z)}\quad (z\in\mathbb{C}\setminus[-1,0[\,).  
\end{equation}

For $u\in]-1,0]$, the expression (19) with $z=u$ holds. Inserting the factorization relation (I, 27), we deduce the relation
\begin{equation}
H(a,u)=\frac{k}{\sqrt{1-a}}\frac{1}{1+ku}\frac{1}{(\gamma T)(a,u)}\frac{1}{N(a,-u)}\;\; (u\in]-1,0]).  
\end{equation}

When compared to Eq. (19), this expression is of poor interest since a Cauchy principal value is again involved. The useful relations for the calculation of $H(a,z)$ are thus (25) in $\mathbb{C}\setminus[-1,0[$, (19) in $]-1,0]$, and (24) at -1.

To our knowledge, the above relations stand as the first {\it complete} solution of the singular $H$-equation. Four previous attemps following the same approach are those by:

- Sobolev$^{[5]}$, in which expression (20) can be found apart from a multiplicative constant, since the question of uniqueness was not understood in 1949. This solution is used again, with no modification, in the first treatise by Sobolev,$^{[6]}$

- Busbridge$^{[7]}$ and Fox$^{[8]}$, who stop at Eq. (10) [with $c(u)=1$], the authors being not aware of Eqs. (I, 32) (with $z = +1/k$) and (I, 33),

- Mullikin$^{[3]}$ and Carlstedt and Mullikin,$^{[9]}$ who clarified the question of uniqueness of the solution and reached formulae (19), (20) and (25) from a technique of analytic continuation. Nowadays, the latter method seems somewhat artificial, and can been avoided by utilizing Eqs. (I, 32) (with $z = +1/k$) and (I, 33). Moreover, the domain of validity of the derived expressions might be stated. These two papers remain however as milestones in the subject, and the source of later works on the singular integral equations of transport theory.

In Ref. [10], the link between the functions $N$ (referred to as $X$) and $H$ is established by using invariant imbedding techniques. Formula (29) in the Sec. 6.2 of this book coincides with our Eq. (25). It is incompatible with the fact that $N(a,+1)<+\infty$, which is suggested on p. 80 of Ref. [10], otherwise $N(a,-1)= \infty$ from Eqs. (I, 26)-(I, 27), and $H(a,+1) = 0$ from Eq. (25). In fact, we have seen in I that the $N$-function is divergent like $T$ or $1/\gamma$ in a neighborhood of +1, which removes this paradox.

\subsection{\label{Subsec. 3.B} Functional relations satisfied by $H$}
Since these relations are well-known (see, e.g., Ref. [11] and references therein), they are not reviewed in detail in the present subsection. Our objective is rather to derive the most important ones, essentially the factorization relation and the non-linear $H$-equation, solely from the definition (17)-(18) for $H$. Also, we intend to extend them to the whole complex plane, since the $H$-function can be defined everywhere in $\mathbb{C}\setminus\lbrace-1/k,-1/K\rbrace$. It can be achieved by exploiting the $N$-function properties from I and the relations connecting the $N$-function to the $H$-function as given in Subsec. \ref{Subsec. 3.A}. Three groups of relations relative to the $N$-function were established in I: the factorization relations (I, 26)-(I, 27), and the integral equations satisfied by $N$ (I, 28)-(I, 31) and $1/N$ (I, 32)-(I, 33). The following three groups of equations for the $H$-function correspond to them:

\noindent
{\it The factorization relation for $H$}:

It can be deduced from the factorization relations (I, 26), (I, 27) and (I, 36) with the help of Eqs. (19)-(20) or (25)-(26). One obtains the unique equation
\begin{equation}
T(a,z)H(a,z)H(a,-z)=1,  
\end{equation}
valid for any complex $z$ except $\pm1/k$, $\pm1$ and $\pm1/K$, which are the zeros of the functions appearing in the left-hand side. It is easy to deduce from this relation the behavior of the $H$-function in a neighborhood of its singular points located at $-1/k$, $-1$ and $-1/K$, considering that this function is regular at the points of opposite abscissae.$^{[11]}$

\noindent
{\it The integral equations satisfied by $H$}:

The integral equations (I, 28)-(I, 31) satisfied by the $N$-function lead to the singular $H$-equation (17) we started from. Putting $z \rightarrow \infty$ in this equation and using $T(a,\infty) = 1-a$ and $H(a,\infty) = 1/\sqrt{1-a}$, we derive the zero order moment of $H$
\begin{equation}
\alpha _{0}(a)=\int_{0}^{1}H(a,v)dv=\frac{2}{a}[1-\sqrt{1-a}],  
\end{equation}
which leads to
\begin{eqnarray}
T(a,z)H(a,z)&=&1+\frac{a}{2}z\int_{0}^{1}H(a,v)\frac{dv}{v-z},\\
&=&\sqrt{1-a}+\frac{a}{2}\int_{0}^{1}H(a,v)\frac{vdv}{v-z}. 
\end{eqnarray}

These integral equations are linear, but singular over $[0,1]$. Non-linear but regular integral equations over $[0,1]$ can be derived thanks to the factorization relation (27). They are the classical $H$-equation and its alternative form
\begin{eqnarray}
\frac{1}{H(a,z)}&=&1-\frac{a}{2}z\int_{0}^{1}H(a,v)\frac{dv}{v+z},\\
&=&\sqrt{1-a}+\frac{a}{2}\int_{0}^{1}H(a,v)\frac{vdv}{v+z}. 
\end{eqnarray}

The well-known Eqs. (29)-(32) are valid in $\mathbb{C}\setminus\lbrace-1/k,-1,-1/K\rbrace$; they play a central role in the standard (Chandrasekhar's) study of the $H$-function.$^{[12]}$

\noindent
{\it The integral equations satisfied by $1/H$}:

Less widely known than the previous ones, these equations follow from the integral equations (I, 32)-(I, 33) for the function $1/N$. To derive them, we need the factorization relation (27) and the relation 
\begin{equation}
\frac{a}{2}\int_{0}^{1}(g/H)(a,v)dv=\frac{1}{\sqrt{1-a}}-1-\frac{R(a,k)}{H(a,+1/k)},
\end{equation}
resulting from Eqs. (I, 32) (with $z=+1/k$) and (22). The $g$-function is defined by Eq. (43) below.

We obtain for $z\in\mathbb{C}\setminus\lbrace [-1,0[\cup\lbrace-1/k\rbrace\rbrace$
\begin{eqnarray}
H(a,z)&=&\frac{1}{1+kz}[1+\frac{kz}{\sqrt{1-a}}]\nonumber\\
&+&\frac{1}{1+kz}\frac{a}{2}z\int_{0}^{1}(g/H)(a,v)(1-kv)\frac{dv}{v+z},\\ 
&=&1\!+\!\frac{R(a,k)}{H(a,+\!1\!/k)}\frac{kz}{1\!+\!kz}+\frac{a}{2}z\int_{0}^{1}(g/H)(a,v)\frac{dv}{v\!+\!z},\\ 
&=&\frac{1}{\sqrt{1-a}}-\frac{R(a,k)}{H(a,+1/k)}\frac{1}{1+kz}\nonumber\\
&-&\frac{a}{2}\int_{0}^{1}(g/H)(a,v)\frac{vdv}{v+z},
\end{eqnarray}
for $u\in ]-1,0]$       
\begin{eqnarray}
(gT/H)(a,-u)&=&\frac{1}{1+ku}[1+\frac{ku}{\sqrt{1-a}}]\nonumber\\
&+&\frac{1}{1+ku}\frac{a}{2}u\overset{1}{\underset{0}{f}}(g/H)(a,v)(1-kv)\frac{dv}{v+u},\\ 
&=&1+\frac{R(a,k)}{H(a,+1/k)}\frac{ku}{1+ku}\nonumber\\
&+&\frac{a}{2}u\overset{1}{\underset{0}{f}}(g/H)(a,v)\frac{dv}{v+u},\\ 
&=&\frac{1}{\sqrt{1-a}}-\frac{R(a,k)}{H(a,+1/k)}\frac{1}{1+ku}\nonumber\\
&-&\frac{a}{2}\overset{1}{\underset{0}{f}}(g/H)(a,v)\frac{vdv}{v+u}, 
\end{eqnarray}
and at $u=-1$
\begin{equation}
\frac{a}{2}\int_{0}^{1}(g/H)(a,v)(1-kv)\frac{dv}{v-1}=1-\frac{k}{\sqrt{1-a}},
\end{equation}
\begin{equation}
\frac{a}{2}\int_{0}^{1}(g/H)(a,v)\frac{dv}{v-1}=1-\frac{k(1+k)}{k^2+a-1}\frac{1}{H(a,+1/k)},
\end{equation}
\begin{equation}
\frac{a}{2}\int_{0}^{1}(g/H)(a,v)\frac{vdv}{v-1}=\frac{1}{\sqrt{1-a}}-\frac {1+k}{k^2+a-1}\frac{1}{H(a,+1/k)}, 
\end{equation}
where the $g$-function is defined for $v\in[0,+1[$ by
\begin{equation}
g(a,v)=\gamma^2(a,v)=\frac{1}{T^2(a,v)+(\pi\frac{a}{2}v)^2}. 
\end{equation}

These equations can be extended by continuity at $u=-1/k$ by
\begin{eqnarray}
\frac{a}{2}\int_{0}^{1}(g/H)(a,v)\frac{dv}{1-kv}=-1-\frac{R(a,k)}{H(a,+1/k)}\nonumber\\
\times[1-\frac{ak^2}{(k^2\!+\!a\!-\!1)(1\!-\!k^2)}+\frac{1}{k}&&\!\!\!\frac{H'(a,\!+1/k)}{H(a,\!+1/k)}], 
\end{eqnarray}
where $H'(a,+1/k)$ is the derivative of the $H$-function at $+1/k$.

Equations in this series made brief appearances in the litera\-ture, specially Eq. (35).$^{[7,13-17]}$ Actually, they are of great importance when solving the radiative transfer equation in a semi-infinite atmosphere illuminated by any source distribution (internal or external). They yield the radiation field leaving the atmosphere from its expression within the medium.$^{[13]}$ Consider for example the classical albedo problem, dealing with the diffuse reflection of light in a half-space irradiated by parallel rays at an angle $\arccos(u)$ $(0<u<1)$ from the inner normal to the surface. The mean intensity of the total field at level $\tau\geq0$ is $(1/4)B(a,\tau,u)$, where $B(a,\tau,u)$ is solution to the integral equation
\begin{equation}
B(a,\tau,u)=\exp(-\tau/u)+\frac{a}{2}\int_{0}^{+\infty}E_{1}(\vert\tau-t\vert)B(a,t,u)dt. 
\end{equation}
Here, $E_1$ is the first exponential integral function as defined by $E_1(\tau)= \int_{0}^{1}\exp(-\tau/u)du/u$ ($\tau>0$).

The surface value of the $B$-function was calculated for the first time by Ambartsumian, who found $B(a,0,u)=H(a,u)\,.^{[18]}$ The solution within the medium is$^{[19]}$
\begin{eqnarray}
B(a,\tau,u)&=&\frac{R(a,k)}{H(a,+1/k)}\exp(-k\tau)H(a,u)\frac{ku}{1-ku}\nonumber\\
&+&(gT)(a,u)\exp(-\tau/u)\nonumber\\
&+&H(a,u)\frac{a}{2}u\overset{1}{\underset{0}{f}}(g/H)(a,v)\exp(-\tau/v)\frac{dv}{v-u}.
\end{eqnarray}

As expected, the $H$-function is retrieved on the $\tau = 0$ layer from this expression and Eq. (38), which illustrates the role played by the latter equation in this particular case.

\section{\label{Sec4} BACK TO THE GENERAL CASE: FOR ANY $c$}
Obviously, the $H$-function is the proper auxiliary function to write the unique solution to Eq. (1) analytic in the right complex half-plane. We are then tempted to express this latter solution, as given by Eqs. (9)-(13), in terms of the $H$-function with the help of relations (19)-(23). Actually, it is possible to solve Eq. (1) following a more direct approach, leading to its solution as a unique functional of $c(z)$ and $H(a,z)$. This relation is valid in the whole complex plane, and it once again yields relations (9)-(13) (expressed in terms of $H$) when the $z$-variable covers their validity domain. It is established hereafter from Eq. (1) through a residue calculation, taking into account the properties ({\it i})-({\it iii}) in Sec. \ref{Sec1} of the free term $c(z)$.

Replace $z\ne0$ by $-1/z'$ in Eq. (1), multiply the resulting equation by $(1/2i\pi)H(a,1/z')/(1+zz')$ and integrate with respect to $z'$ along the imaginary axis in the sense of the principal value at infinity, i.e., by writing $\int_{-i\infty}^{+i\infty}=\lim_{X\to+\infty} \int_{-iX}^{+iX}$. We obtain
\begin{eqnarray}
\lefteqn{\frac{1}{2i\pi}\int_{-i\infty}^{+i\infty}H(a,1/z')T(a,1/z')f(-1/z'){\frac{dz'}{1+zz'}}^{(\alpha)}}\nonumber\\&&\qquad+\frac{1}{2i\pi}\int_{-i\infty}^{+i\infty}H(a,1/z')\frac{a}{2}\int_{0}^{1}f(v)\frac{dv}{1+vz'}{\frac{dz'}{1+zz'}}^{(\beta)}\nonumber\\&& \qquad\qquad\qquad=\frac{1}{2i\pi}\int_{-i\infty}^{+i\infty}H(a,1/z')c(-1/z')\frac{dz'}{1+zz'}. 
\end{eqnarray}

The ($\alpha$)-term on the left-hand side can be calculated by replacing $H(a,1/z')T(a,1/z')$ by $1/H(a,-1/z')$, observing that the function $z'\rightarrow f(-1/z')/H(a,-1/z')$ is analytic in the left complex half-plane. Applying the theorem of residues to an obvious contour in the left complex half-plane, we find for all $z\in\mathbb{C}^*$    
\begin{equation}
(\alpha)=\frac{1}{z}\lbrace{\rm Y}[\Re(z)]\frac{f(z)}{H(a,z)}-\frac{1}{2}f(+0)\rbrace, 
\end{equation}
where Y is the unit step (or Heaviside) function extended by 1/2 at 0 [${\rm Y}(x)$ is 0 for $x<0$, 1/2 for $x=0$, and 1 for $x>0$]. Owing to this extension, (48) is still valid on the imaginary axis without zero (denoted $i\mathbb{R}^*$), the integral in the ($\alpha$)-term of Eq. (47) becoming a Cauchy principal value.

The ($\beta$)-term on the left-hand side of Eq. (47) is calculated by inverting the order of integration, writing
\[
\frac{1}{1+vz'}\frac{1}{1+zz'}=[\frac{1}{z'+1/v}-\frac{1}{z'+1/z}]\frac{1}{v-z}\,.
\]
It follows from the residue theorem, applied to a contour in the right complex half-plane, that
\[
\frac{1}{2i\pi}\!\int_{-i\infty}^{+i\infty}\!H(a,1/z')\frac{dz'}{(1\!+\!vz')(1\!+\!zz')}={\rm Y}[-\Re(z)]H(a,-z)\frac{1}{v\!-\!z}, 
\]
even when $z\in i\mathbb{R}^*$ since Y(0) = 1/2. We obtain
\begin{eqnarray}
(\beta)&=&{\rm Y}[-\Re(z)]H(a,-z)\frac{a}{2}\int_{0}^{1}f(v)\frac{dv}{v-z},\nonumber\\ 
&=&\frac{1}{z}{\rm Y}[-\Re(z)]H(a,-z)[T(a,z)f(z)-c(z)]\,. 
\end{eqnarray}

From Eqs. (48), (11), (49) and the factorization relation (27), Eq. (47) leads to the following solution to Eq. (1), analytic in the right complex half-plane:
\begin{eqnarray}
\lefteqn{f(z)=H(a,z)\left\lbrace\frac{1}{2}c(+0)+{\rm Y}[-\Re(z)]H(a,-z)c(z)\right.}\nonumber\\&&\left.\qquad\qquad\qquad+\frac{z}{2i\pi}\int_{-i\infty}^{+i\infty}H(a,1/z')c(-1/z')\frac{dz'}{1+zz'}\right\rbrace. 
\end{eqnarray}

This expression is valid at any $z\in\mathbb{C}$ where Eq. (1) has a solution. In particular, it is satisfied on the imaginary axis provided that ${\rm Y}(0)=1/2$. When $c(z)=1$, we retrieve $f(z)=H(a,z)$ since the integral within the brackets is $1/2-{\rm Y}[-\Re(z)]H(a,-z)$.

This concise expression is general, and the details of the usual mathematical developements for solving problems in the class of (1)--that is reducing them to a Hilbert problem--are no longer necessary. The conciseness of the expression (50) will be turned to account in the forthcoming analytical developments regarding the $X$- and $Y$-functions for finite media (third paper in this series). 

To numerically evaluate the $f$-function, it is necessary to transform the integral in Eq. (50) with the help of the residue theorem. We obtain different expressions for the solution, depending on the location of the argument in the complex plane. The calculation is classical and leads to the following results:

\noindent
- if $z\in\mathbb{C}\setminus\lbrace[0,1]\cup\lbrace +1/k\rbrace\rbrace$
\begin{eqnarray}
\lefteqn{f(z)=H(a,z)\left[\frac{R(a,k)}{H(a,+1/k)}c(+1/k)\frac{kz}{1-kz}+H(a,-z)c(z)\right.}\nonumber\\
&&\left.\qquad\qquad\qquad\qquad\qquad+\frac{a}{2}\,z\int_{0}^{1}(g/H)(a,v)c(v)\frac{dv}{v-z}\right],
\end{eqnarray}
- if $u \in]0,1[$
\begin{eqnarray}
\lefteqn{f(u)=H(a,u)\left[\frac{R(a,k)}{H(a,\!+1\!/k)}c(+1\!/k)\frac{ku}{1\!-\!ku}+(gT/H)(a,u)c(u)\right.}\nonumber\\
&&\left.\qquad\qquad\qquad\qquad+\frac{a}{2}\,u\overset{+1}{\underset{0}{f}}(g/H)(a,v)c(v)\frac{dv}{v-u}\right],
\end{eqnarray}
- at $u=0$
\begin{equation}
f(+0)=c(+0)<+\infty,
\end{equation}
- at $u=+1$
\begin{eqnarray}
\lefteqn{f(+1)=H(a,+1)\left[\frac{R(a,k)}{H(a,+1/k)}c(+1/k)\frac{k}{1-k}\right.}\nonumber\\
&&\left.\qquad\qquad\qquad\qquad\quad+\frac{a}{2}\int_{0}^{1}(g/H)(a,v)c(v)\frac {dv}{v-1}\right], 
\end{eqnarray}
- and at $u=+1/k$
\begin{eqnarray}
\lefteqn{f(+1/k)=H(a,+1/k)\left \lbrace \frac{R(a,k)}{H(a,+1/k)}\frac{1}{k}c'(+1/k)+c(+1/k)\right.}\nonumber\\
&&\left.\qquad\qquad\qquad-\frac{a}{2}\int_{0}^{1}(g/H)(a,v)[c(v)-c(+1\!/k)]\frac{dv}{1\!-\!kv}\right \rbrace. 
\end{eqnarray}

These relations could have been derived by expressing the $N$-function in terms of the $H$-function in Eqs. (9)-(13). We find again $f = H$ by putting $c = 1$ in them, due to the integral equations (35), (38) and (41).

\section{CONCLUSION}
We used the theory of Cauchy integral equations to the solution of equations in the form (1) which appear in the linear transport theory assuming isotropic scattering. We felt necessary a revisited description of the solution to this type of equation for the following reasons:

\noindent
1) The pionneering studies of the sixties may be presented more simply than formerly by introducing the values on the cut of any function defined by a Cauchy-type integral.

\noindent
2) In the fifties and sixties, the factorization relation (27) and the non-linear $H$-equation (31) were well-trodden starting points for the study of the $H$-function. However, no completely self-consistent approach was available starting from the linear $H$-equation (17) as constrained by Eq. (18). For instance the papers by Mullikin et al.$^{[3,9]}$ presuppose the factorization relation (27) usually established from the Wiener-Hopf method. In Case and Zweifel's mono\-graph,$^{[10]}$ the technique of invariant imbedding is used to establish the relation connecting the $H$-function to the $N$-function.

\noindent
3) The interest in studying the $H$-function from its definition (17)-(18) is not only of pedagogical essence: this approach can be enlarged to the study of finite media. The difficulty in solving Eq. (1) in a finite slab originates in the dependence of the free term $c(z)$ on the unknown function $f(z)$. It will be dealt with in the third paper of this series with the help of formula (50), since the solution has to be analytic in the right complex half-plane (and in the left one as well).

\end{document}